\documentclass[authoryear,12pt,3p,letter]{jowarticle}
\usepackage{graphicx}
\usepackage{amsthm,amsmath}
\usepackage{amssymb}
\usepackage{amsfonts}
\usepackage{natbib}
\usepackage{setspace}
\usepackage[all]{xy}
\usepackage{enumitem}
\usepackage{titlesec}
\usepackage{mathrsfs}
\makeatletter
\renewcommand\section{\@startsection{section}{1}{\z@}{-3.25ex plus -1ex minus -.2ex}{1.5ex plus .2ex}{\normalsize\bf}}
\renewcommand\subsection{\@startsection{subsection}{2}{\z@}{-3.25ex plus -1ex minus -.2ex}{1.5ex plus .2ex}{\normalsize\bf}}
\renewcommand\subsubsection{\@startsection{subsubsection}{3}{\z@}{-3.25ex plus -1ex minus -.2ex}{1.5ex plus .2ex}{\normalsize\bf}}
\makeatother

\newtheorem{innercustomgeneric}{\customgenericname}
\providecommand{\customgenericname}{}
\newcommand{\newcustomtheorem}[2]{%
  \newenvironment{#1}[1]
  {%
   \renewcommand\customgenericname{#2}%
   \renewcommand\theinnercustomgeneric{##1}%
   \innercustomgeneric
  }
  {\endinnercustomgeneric}
}

\newtheorem{thm}{Theorem}
\newtheorem{cor}[thm]{Corollary}

\newtheorem{prop}[thm]{Proposition}

\newcustomtheorem{prop*}{Proposition}
\newcustomtheorem{cor*}{Corollary}
\begin{document}
\begin{frontmatter}
\title{Geometry and Motion in General Relativity}
\author{James Owen Weatherall}\ead{weatherj@uci.edu}
\address{Department of Logic and Philosophy of Science\\ University of California, Irvine}
\begin{abstract}A classic problem in general relativity, long studied by both physicists and philosophers of physics, concerns whether the geodesic principle may be derived from other principles of the theory, or must be posited independently.  In a recent paper [Geroch \& Weatherall, ``The Motion of Small Bodies in Space-Time'', \textit{Comm. Math. Phys.} (forthcoming)], Bob Geroch and I have introduced a new approach to this problem, based on a notion we call ``tracking''. In the present paper, I situate the main results of that paper with respect to two other, related approaches, and then make some preliminary remarks on the interpretational significance of the new approach.  My main suggestion is that ``tracking'' provides the resources for eliminating ``point particles''---a problematic notion in general relativity---from the geodesic principle altogether.\end{abstract}
\end{frontmatter}
                                         % Activate to display a given date or no date
\doublespacing
\section{Introduction}\label{introduction}

There is a deep link in general relativity between, on the one hand, the geometry of spacetime, and on the other hand, the motion of small bodies.  Spacetime in the theory is represented by a smooth manifold $M$ endowed with a smooth metric $g_{ab}$; this metric (and its associated Levi-Civita derivative operator, $\nabla$) determine a class of \emph{timelike geodesics}, which are the curves of ``locally extremal'' length.  These curves have special physical significance in the theory: they are the possible trajectories of free massive test point particles.  Thus we have an identification between a class of geometrically privileged curves and a class of physically privileged trajectories in general relativity.

This link is sometimes called the \emph{geodesic principle}; it is analogous to Newton's first law of motion.  Because of its centrality to the interpretation of spacetime geometry in general relativity, this principle has received a great deal of attention from both physicists and philosophers of physics, going back at least to \citet{Einstein+Grommer}.\footnote{\label{literature} For a recent review of the physics literature on this subject, see \citet{Poisson}; for other recent work, see \citet{Asada}, \citet{Wald+Gralla}, and the contributions to \citet{Puetzfeld}.  For the recent philosophical literature, which generally stems from a discussion of the geodesic principle by \citet{Brown}, see \citet{MalamentGP}, \citet{Tamir}, \citet{Sus}, \citet{Samaroo}, \citet{LehmkuhlCareful, LehmkuhlHybrid}, and \citet{WeatherallSGP,WeatherallLehmkuhl,WeatherallConservation}.}   One issue of particular significance concerns whether the geodesic principle is an independent postulate, or if, instead, it should be understood as a consequence of other principles of the theory.  This question is known as the ``problem of motion'' in general relativity.  In fact, it is widely recognized that the geodesic principle \emph{is}, in some sense, a consequence of the rest of the theory.  But articulating this sense in a satisfactory way turns out to be remarkably subtle.  Over the last century, dozens of different attempts have been made to capture, in a mathematically precise and physically perspicuous way, the sense in which the geodesic principle is a theorem of general relativity.

In a recent paper, Bob Geroch and I have introduced a new approach to the problem of motion \citep{Geroch+Weatherall}.  The main theorem of that paper, Theorem \ref{tt} below, captures a sense in which generic small bodies in general relativity follow timelike geodesics (and light rays follow null geodesics).  This theorem has a number of virtues over other approaches, at least some of which are salient to recent discussions in the philosophy of physics literature concerning the relationship between spacetime geometry and the dynamics of matter.  My goal here is to present the main results of that earlier paper in a way that emphasizes some of these relative virtues, and then make some preliminary remarks on the interpretational significance of the results.  The main suggestion---which is only implicit in the earlier paper---will be that the methods used in stating and proving this theorem provide the resources for eliminating ``point particles''---a problematic notion in general relativity---from the geodesic principle altogether.  Instead, I will argue that one can capture the substance of the link between geometry and motion directly as an assertion about the solutions to the field equations governing realistic, extended matter.

The remainder of the paper will proceed as follows.  In the next section, I will describe two well-known approaches to the problem of motion, and discuss some of their shortcomings.  In the following section I will present the main results from \citet{Geroch+Weatherall} and explain how they combine the virtues of these two other approaches while avoiding their problems.  I will conclude by discussing how one might re-think the geodesic principle in light of these results.  I emphasize that I will not attempt to reproduce the discussion in \citet{Geroch+Weatherall}, and I direct the reader there for many of the mathematical details and for proofs of propositions.  Rather, my goal is to give a different, complementary presentation of (some of) that material, with an emphasis on its motivation, what makes it distinctive, and some of the reasons why it might be of interest to philosophers.

\section{Two Approaches, and Their Discontents}\label{sec:two}

In what follows, fix a relativistic spacetime, $(M,g_{ab})$.\footnote{Although this is meant to be a relatively gentle introduction, I take for granted the basic mathematics of general relativity; for relevant background, see, for instance, \citet{Wald} or \citet{MalamentGR}, both of whom use essentially the same notation as I do.}  The geodesic principle states that free massive test point particles traverse timelike geodesics of spacetime.  In this section, I present two widely discussed approaches to capturing the geodesic principle as a theorem of general relativity, and describe reasons why one might be dissatisfied with each of them.  I do not mean to claim that these are the \emph{only} two approaches in the literature---to the contrary, there are many approaches out there.\footnote{Once again, see the references in note \ref{literature}.  One approach in particular that has been widely influential, but which I do not discuss at all, is the method of matched asymptotic expansion, as developed, for instance, by \citet{Death}, \citet{Thorne+Hartle}, \citet{Mino+etal}, and \citet{Wald+Gralla}.}  But these two are distinguished by the fact that they yield precise mathematical theorems that are strong and simple, and do not rely on physical arguments that call into question the generality of the results.\footnote{\label{math} There is a certain trade-off between, on the one hand, strength and simplicity, and on the other hand, information relevant in special cases, such as possible deviations from geodesic motion that might arise from finite body effects ``on the way to the limit''.  Compare, for instance, \citet{Thorne+Hartle} or \citet{Wald+Gralla}, who describe, in the presence of additional (strong) assumptions, higher order ``corrections'' to geodesic motion for finite bodies, with the results to be described here, which might be understand to characterize (without these strong assumptions) the universal limiting, or order zero, behavior of small bodies.  My perspective is that for foundational purposes, the more general and precise results are of greater value, though this is not necessarily true for other purposes, such as studying binary black holes.  On the other hand, see footnotes \ref{spin1} and \ref{spin2} for ways in which the perspective taken here may bear fruitfully on widely accepted results from other approaches.}

To begin, however, let me comment on (part of) why formulating such theorems requires care.  Basically, the difficulty comes down to the fact that ``free massive test point particles'' are not particularly natural objects in general relativity.  For reasons I discuss below, one usually considers extended matter, represented by smooth fields of various kinds.  In principle, one would like to say that a point particle is an idealization of a ``small body'', and so one would like to associate small extended bodies with curves that they ``traverse''.  In special relativity, as in Newtonian mechanics, there is no difficulty in doing so: one can identify, with any extended body, a unique ``center of mass'' trajectory, reflecting the ``average'' motion of the body, and then argue that that trajectory must be a timelike geodesic.\footnote{\citet{Geroch+Jang} give a compact treatment of the situation in special relativity.  For further discussion of the situation regarding theorems of the present sort in Newtonian gravitation, see \citet{WeatherallMBNT,WeatherallSGP,WeatherallLehmkuhl}.  I will not discuss these results further in the present paper.}

But in curved spacetime, analogous constructions are apparently not possible.  In that context, although there are of course many curves that lie within the worldtube of any given (extended) body, it is not clear that any of them captures the overall motion of the body---and in general, there need not be \emph{any} geodesic lying within the (convex hull) of the worldtube of a body, even in the absence of external forces.  This suggests that, whatever else is the case, the geodesic principle should only hold in the limit as the radius of a body goes to zero; for extended matter in curved spacetime, it is hard to identify even a candidate assertion that captures the idea that bodies move on geodesics.

\subsection{Distributions}

One way to overcome these challenges is to give up on representing bodies with smooth fields, and instead to consider point particles represented as \emph{distributions}---basically, generalized functions, such as $\delta$ functions and their derivatives---that are supported only on curves.  Roughly, a distribution is a map from a space of \emph{test fields}, which are usually smooth fields of compact support, to the real numbers that is continuous in a suitable sense.\footnote{More precisely, we take test fields to be \emph{densities} of weight 1; see \citet[Appendix A]{Geroch+Weatherall}.}  Although they are not smooth functions, one can generally manipulate them as if they were, for instance by taking their derivatives.\footnote{We take derivatives by analogy with integration by parts.  Fix a manifold $M$, a derivative operator $\nabla$ on $M$, and a distribution $\mathbf{X}$ on $M$.  Then $\nabla_a\mathbf{X}$ is that distribution whose action on a smooth test field $\alpha^a$ is given by $\nabla_a\mathbf{X}\{\alpha^a\} = -\mathbf{X}\{\nabla_a \alpha^a\}$. For background on distributions, including tensor distributions, see \citet[Appendix A]{Geroch+Weatherall}, \citet{distributionGR}, or \citet{Steinbauer+Vickers}.  The details of the theory of distributions do not particularly matter for the arguments that follow.}  As first observed by \citet{Matthisson}, and developed by \citet{Souriau}, \citet{Sternberg+Guillemin}, and others, this approach leads to a very short argument for geodesic motion.

The argument goes as follows.  Suppose one is given a symmetric distribution ${\mathbf T}^{ab}$ supported on a timelike curve $\gamma$ in $(M,g_{ab})$.  We might take this distribution to represent the energy-momentum of a small body---or better, a point particle, since it has no spatial extension.  Now suppose this distribution is order zero and divergence-free, where by order zero we mean the action of $\mathbf{T}^{ab}$ may be extended from smooth fields of compact support to merely continuous fields (which means, roughly, that the value ${\mathbf T}^{ab}$ yields when acting on a test field $\alpha_{ab}$ depends only on the value of $\alpha_{ab}$ at each point, and not its derivatives), and by divergence-free, we mean that $\nabla_a {\mathbf T}^{ab} = \mathbf{0}$.\footnote{That is, ${\mathbf T}^{ab}$ vanishes on all test fields of the form $\nabla_a \alpha_b$.}  It follows, by a short calculation, that ${\mathbf T}^{ab} = m{\boldsymbol\delta}_{\gamma} u^a u^b$, where $m$ is a number, ${\boldsymbol \delta}_{\gamma}$ is the delta distribution supported on $\gamma$, and $u^a$ is the unit tangent to $\gamma$.  It also follows that $\gamma$ is a (timelike) geodesic.

This approach has some obvious advantages.  The argument just given is mathematically very simple.  It is also easy to generalize to forces.  For instance, still assuming ${\bf T}^{ab}$ order zero, a body with timelike worldline $\gamma$, subject to an arbitrary force ${\bf f}^a = \nabla_b {\bf T}^{ab}$, can be described by an energy-momentum ${\bf T}^{ab} = {\boldsymbol\mu} u^a u^b$ satisfying
\begin{align*}
{\boldsymbol \mu}u^n\nabla_n u^a &= q^a{}_b{\bf f}^b\\
\nabla_b({\boldsymbol \mu}u^b) &= -{\bf f}^b u_b
\end{align*}
where ${\boldsymbol\mu}$ is an order zero distribution supported on $\gamma$.  The first of these equations asserts precisely that $\mathbf{F}=m\mathbf{a}$; and the second is a ``continuity'' equation describing the possibility of transfer of mass between different bodies.

Likewise, fix a background electromagnetic field $F_{ab}$.  Represent a charged body by an energy-momentum distribution ${\bf T}^{ab}$ supported on a timelike curve $\gamma$ and an (order zero) charge current density ${\bf J}^a$, also supported on $\gamma$.  Assume $\nabla_a \mathbf{J}^a=0$ and ${\bf f}^a = F^a{}_b {\bf J}^b$.  Then ${\bf J}^a = e{\boldsymbol \delta}_{\gamma}u^a$, ${\bf T}^{ab} = m {\boldsymbol \delta}_{\gamma} u^a u^b$, and $\gamma$ is a $e/m$ Lorentz force curve, that is, $u^n\nabla_n u^a = e/m F^a{}_bu^b$.  Moreover, one can solve for the general case, where ${\bf J}^a$ is order one (the highest order compatible with ${\bf T}^{ab}$ being order zero); one finds contributions to the motion arising from electric and magnetic dipoles.

So distributions do not merely capture the idea of ``free'' motion; they also allow us to derive general claims about particle motion, including the Lorentz force law.  But despite this simplicity and power, the situations concerning distributions is not entirely satisfactory.

One concern is immediate.  We assumed, from the beginning, that the distribution ${\bf T}^{ab}$ representing the energy-momentum of a point particle is order zero.  Without this assumption, none of the arguments above go through, and indeed, one can find divergence-free distributions on any curve at all.\footnote{This is the distributional analog to the result proved in \citet{MalamentGP}.}  But why assume this?

In fact, the restriction to order zero distributions can be justified by the following argument.  Let us say that a smooth test field $t_{ab}$ satisfies the \emph{dual energy condition} at a point $p$ if $t_{ab}$ can be written as a sum of symmetrized outer products of co-oriented causal covectors.  The fields satisfying this condition at a point are precisely the ones that are ``dual'' to tensors $T^{ab}$ satisfying the (standard) dominant energy condition, which states that given any pair of co-oriented causal vectors $\eta^a$ and $\xi^a$, $T^{ab}\xi_a\eta_b \geq 0$.  We then say that a symmetric distribution ${\bf T}^{ab}$ satisfies the \emph{dominant energy condition} if, for every test field $t_{ab}$ satisfying the dual energy condition, ${\bf T}^{ab}\{t_{ab}\} \geq 0$.  Note that this is a straightforward extension of the dominant energy condition from tensors at a point (and smooth tensor fields) to distributions.

We then get the following result.
\begin{prop} Let ${\bf T}^{ab}$ be a symmetric distribution satisfying the dominant energy condition.  Then ${\bf T}^{ab}$ is order zero.
\end{prop}
\noindent Thus, insofar as one expects matter to satisfy the dominant energy condition, it should be represented by distributions that are order zero.\footnote{\label{spin1} Although it is a side issue for present purposes, observe that this result points to a problem with certain approaches to treating the motion of rotating particles that represent ``spin'' by higher order distributions supported on a curve \citep{Papapetrou, Souriau}: such particles are incompatible with the energy condition.  There is good physical reason for this.  For ever smaller bodies to have large angular momentum (per unit mass), their rotational velocity must increase without bound---leading to superluminal velocities, which are incompatible with the energy condition.}

But other concerns are less easily dealt with.  In particular, although distributions seem like a natural way of representing ``point particles'' in general relativity, it is difficult to see how they are related to ``realistic'' matter.  As I noted above, matter in general relativity is usually represented by smooth fields.  These fields are generally solutions to certain systems of partial differential equations, such as Maxwell's equations or the Klein-Gordon equation; each such solution is associated with some (smooth) energy-momentum tensor, via standard formulae.  For standard examples (Maxwell, Klein-Gordon, etc.), energy-momentum tensors are quadratic in field values and/or their derivatives.

But this commonplace observation is a big problem for the distributional approach.  If we consider only smooth solutions to these equations, then the associated energy-momentum tensors will also be smooth, i.e., they will not be distributions supported on a curve.  So the distributional energy-momenta considered above cannot arise in this way.  One might think that this means we should consider distributional solutions to the matter field equations, in which case one could perhaps find solutions that are supported on a curve.  But even if one had such a solution, one could not generally associate an energy-momentum tensor with it.  The reason is that multiplication of distributions is not well-defined.\footnote{There are extensions to the theory of distributions---namely, the theory of Colombeau algebras \citep{Colombeau}---that permit one to multiply distributions.  But these have some undesirable properties, including that multiplication is not uniquely defined for distributions, and it does not reduce to pointwise multiplication for all (continuous) functions, conceived as distributions.}  And so it is not clear how the distributional energy-momenta we have been considering are supposed to arise, or what kind of matter they represent.

A related difficulty arises when we try to understand distributional energy-momenta as sources in Einstein's equation.  In fact, a well-known result due to \citet{Geroch+Traschen} establishes that there are no metrics satisfying certain weak conditions compatible with distributional sources supported on a curve.  Thus it is difficult to evaluate, for instance, backreaction arising from a distributional ${\bf T}^{ab}$.

And so it seems that distributional energy-momenta cannot arise from realistic matter, and they cannot act as sources in Einstein's equation.  So in what sense do distributional ${\bf T}^{ab}$ represent anything physical?  And what bearing do the simple results described above have on the motion of actual bodies?

\subsection{Curve-First}

A second approach to the problem of motion, which has been widely discussed in the philosophical literature, was developed by \citet{Geroch+Jang} and \citet{Ehlers+Geroch}.  On this approach, one begins with a curve $\gamma$ and considers smooth fields $T^{ab}$, satisfying the dominant energy condition, supported in small neighborhoods of the curve.  These fields represent the energy-momenta of small bodies propagating ``near'' the curve $\gamma$.  One then proves the following theorem.
\begin{thm}[Geroch-Jang]
Let $\gamma$ be a smooth, timelike curve in a spacetime $(M,g_{ab})$.  Suppose that, in any neighborhood $O$ of $\gamma$, there exists a smooth, symmetric, divergence-free, and non-vanishing tensor field $T^{ab}$ satisfying the dominant energy condition whose support lies in $O$. Then $\gamma$ is a geodesic.\footnote{Observe that we assume from the start that the curve is timelike; if one wants to \emph{conclude} that the curve must be timelike, a stronger energy condition is required \citep{WeatherallEC}.}
\end{thm}
The interpretation of this result is perhaps not quite as straightforward as the distributional result.  The idea is that the only curves along which arbitrarily small massive bodies (represented by spatially localized $T^{ab}$ fields, satisfying the dominant energy condition) may propagate in the absence of any external forces (captured here by the requirement that the fields be divergence-free) are (timelike) geodesics.  Thus we get a sense in which free massive point particles must follow timelike geodesics.\footnote{For further discussion of the interpretation of this theorem, see \citet{WeatherallSGP,WeatherallLehmkuhl}; I do not wish to belabor here points I already make elsewhere.}

Like the distributional approach, the curve-first approach is also simple.   And since it refers only to smooth $T^{ab}$ fields, its physical interpretation is more transparent.  Moreover, smooth $T^{ab}$ fields may be sources in Einstein's equation, and so this method may be adapted to consider backreaction.  Indeed, there is a strengthening of the Geroch-Jang theorem that captures precisely this:
\begin{thm}[Ehlers-Geroch]
Let $\gamma$ be a smooth, timelike curve in a spacetime $(M,g_{ab})$.  Suppose that, for any (closed) neighborhood $O$ of $\gamma$, and any $C^1[O]$ neighborhood $\hat{O}$ of $g_{ab}$, there exists a Lorentzian metric $\hat{g}\in\hat{O}$ whose Einstein tensor is non-vanishing, which satisfies the dominant energy condition (relative to $g_{ab}$), and whose support lies in $O$. Then $\gamma$ is a geodesic.
\end{thm}
The interpretation of this result is that even if we consider small bodies that ``perturb'' the spacetime metric $g_{ab}$, at least to first order, in the limit as those bodies become small (in mass and spatial extent), they must follow timelike geodesics of $g_{ab}$.

But again, the situation is not totally satisfactory.  One issue is that curve-first results work well for \emph{free} bodies, but it is difficult to see how to generalize them to include forces---including, for instance, the Lorentz force law, which one might have guessed would have a similar status as the geodesic principle.\footnote{\citet{Harte} extend a version of a curve-first approach to treat the Lorentz force law, and also derive leading order ``self-force'' corrections to it.  But the relationship between their arguments are the sort of result envisaged here is the same as the relationship between the \citet{Wald+Gralla} results and, say, the \citet{Geroch+Jang} theorem, which is that they require much stronger assumptions.  (Recall footnote \ref{math}.)}  Recall that in the distributional case, these sorts of generalizations were almost immediate, because the energy condition placed a strong constraint on possible forces and also on (for instance) charge-current densities.  But on the curve-first approach, the energy condition does not seem to place analogous constraints on smooth $T^{ab}$ fields.  It seems some further conditions are needed to recover the equations characterizing forces.

Another concern is that, although curve-first results consider smooth fields, there is still a problem concerning ``realistic'' matter, in the form of solutions to some hyperbolic system.  The issue now has to do with the way in which the limit is taken.  In particular, the Geroch-Jang and Ehlers-Geroch theorems assume matter fields can vanish outside of arbitrary neighborhood of a timelike curve.  But for hyperbolic systems, this is not generally possible: solutions to the Maxwell or Klein-Gordon equations, for instance, tend to spread over time, and there are, in general, no solutions that are supported arbitrarily closely to a curve for all time.  This leads to the following embarassing situation: the geodesic principle theorems do not establish that Maxwell fields follow null geodesics, even in an appropriate high-frequency (optical) limit!\footnote{For instance, in his classic textbook \citet{Wald} describes the Geroch-Jang theorem as capturing the sense in which small bodies follow timelike geodesics, but then does not invoke this result to establish that light rays traverse null geodesics---appealing, instead, to a completely different construction.}

Thus we find that curve-first results, like distributional results, are of limited physical applicability.  In particular, it is not clear how to think of solutions of the field equations that govern real matter in general relativity as somehow realizing the conditions assumed in the limiting procedure for these results.

\section{The Miracle of Tracking}\label{sec:tracking}

We saw in the last section that both the distributional and curve-first approaches have some attractive features---but that neither is fully satisfactory.  In this section, I describe a novel approach to the problem that combines the distributional and curve-first approaches, and does so in a way that allows us to extend both while also clarifying the physical significance of both constructions.\footnote{There is a sense in which \citet{Wald+Gralla} and \citet{Harte} also combine features of both approaches, though their approach is considerably different.  Recall, again, footnote \ref{math}.}  The results here are from \citet{Geroch+Weatherall}; proofs of all propositions, as well as further discussion emphasizing different issues, can be found there.

\subsection{Definition of tracking}

The key concept in this approach is that of \emph{tracking}, which we introduce now.  To begin, fix, once again, a relativistic spacetime $(M,g_{ab})$.  Let us suppose that we are given, on this spacetime, a collection ${\mathcal C}$ of smooth, symmetric fields $T^{ab}$, each satisfying the dominant energy condition.  Although these fields are smooth, each of them is naturally associated with a distribution, the action of which on test fields $x_{ab}$ is defined by $\mathbf{T}^{ab}\{x_{ab}\}=\int_M T^{ab}x_{ab}$.  We will say that this collection \emph{tracks} a timelike curve $\gamma$ if, for every smooth test field ${x}_{ab}$ satisfying the dual energy condition in a neighborhood of $\gamma$ and generic at some point of $\gamma$,\footnote{By ``generic'' at a point $p$, I mean that $x_{ab}$ lies in the interior of the cone of tensors satisfying the dual energy condition at a $p$: that is, for any non-vanishing tensor $T^{ab}$ satisfying the dominant energy condition at $p$, $T^{ab}x_{ab} > 0$.} there is a field $T^{ab}$ in ${\cal C}$ such that $\mathbf{T}^{ab}\{{x_{ab}}\} > 0$.\footnote{Observe the notational convention adopted here: previously we had used boldface for distributions; now we are using bold symbols to refer to the distributions associated with (determined by) smooth fields represented by the same, non-bolded, symbol.}

The rest of this section concerns facts about, and applications of, tracking.  Since this concept is the main idea in what follows, its interpretation desires special attention.  First, observe that because each field $T^{ab}$ in the collection $\mathcal{C}$ satisfies the dominant energy condition, when you contract it, at a point, with a test field that satisfies the dual energy condition there, the result is non-negative.  (Observe that this makes sense, since the fields in $\mathcal{C}$ are ordinary smooth tensor fields; they determine distributions, but we can also consider their action on vectors and covectors at a point.)  This means that when a field in $\mathcal{C}$ acts, as a distribution, on a test field that satisfies the dual energy condition everywhere, then the result is necessarily non-negative (though it may vanish).  (This, recall, is just what it means to say that a distribution satisfies the dominant energy condition.)  We may thus think of any given test field $x_{ab}$, satisfying the dual energy condition everywhere, as giving a standard of ``magnitude'' for $T^{ab}$ in the region where $x_{ab}$ is supported, with different test fields giving different standards.

\begin{figure}[ht]
\center
\includegraphics[height=.25\paperheight]{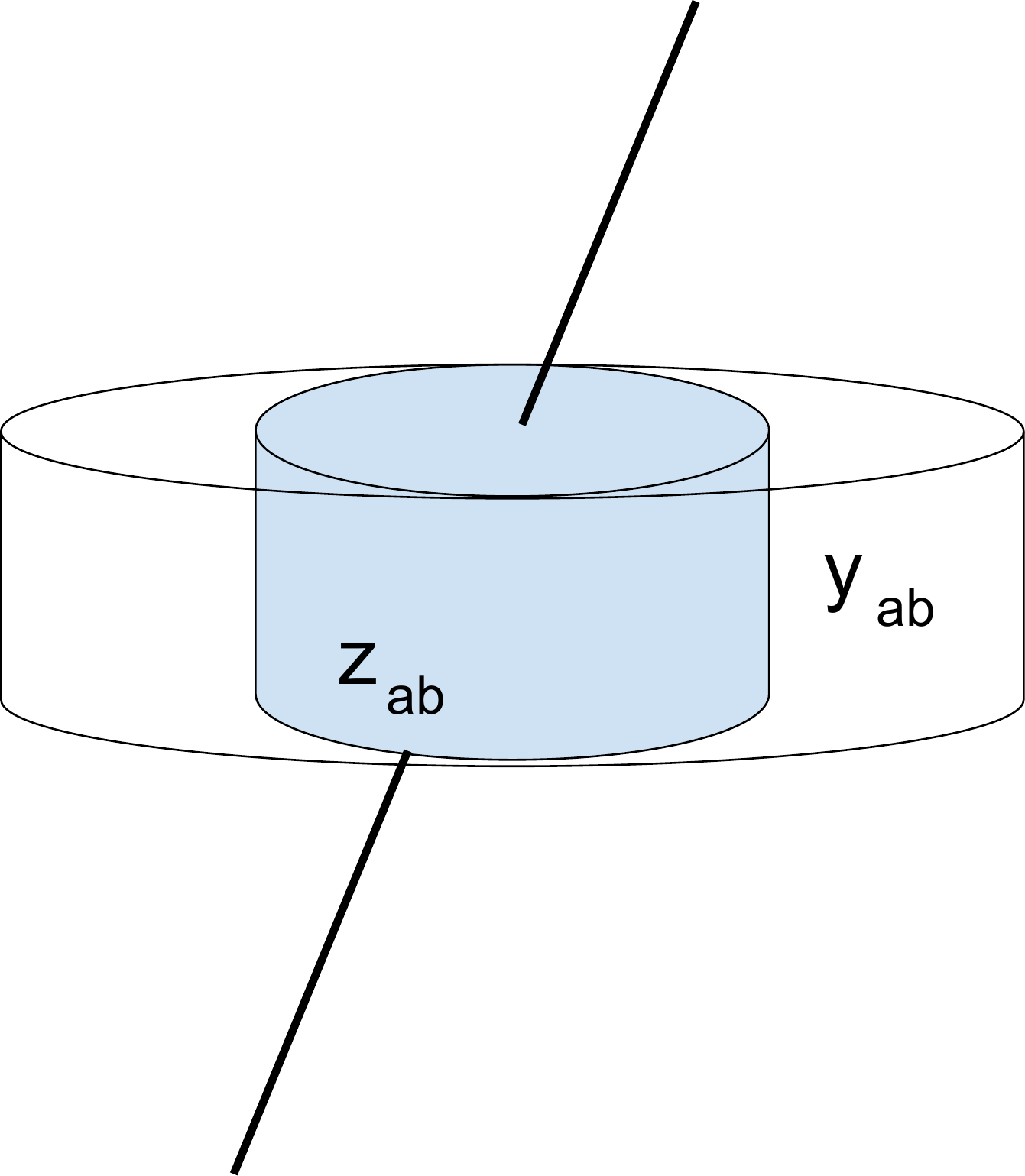}
\caption{Here we depict the basic construction underlying the notion of ``tracking''.  Consider two test fields, $z_{ab}$ and $y_{ab}$, both satisfying the dual energy condition, but where $z_{ab}$ is supported near a curve $\gamma$ and $y_{ab}$ is supported away from $\gamma$.  Then $x_{ab} = z_{ab}-y_{ab}$ satisfies the dominant energy condition near $\gamma$.  If $T^{ab}$ satisfying the dominant energy condition satisfies $T^{ab}\{x_{ab}\} > 0$, then there is ``more'' $T^{ab}$ in the region where $z_{ab}$ is supported than the region where $y_{ab}$ is supported.}
\label{fig:tracking}
\end{figure}

Of course, none of this holds if one acts on test fields that satisfy the dual energy condition only at some points---in that case, fields in $\mathcal{C}$ may or may not yield a non-negative result.  Given a test field $x_{ab}$, however, satisfying the dual energy condition in a region $O$ (and non-vanishing there), one can always construct a field $T^{ab}$, satisfying the dominant energy condition, whose action, as a distribution, on $x_{ab}$ is positive, by ensuring that $T^{ab}$ is sufficiently ``large'' in $O$ and ``small'' in $M-O$ (by the standard of ``large'' and ``small'' given by $x_{ab}$). That is, one can choose $T^{ab}$ so that the part of the integral taken over $O$ dominates, i.e., so that
\[
\int_{O} T^{ab}x_{ab} \geq \left|\int_{M-O} T^{ab}x_{ab}\right|.
\]
Indeed, this interpretation is particularly clear in cases where a test field $x_{ab}$ may be decomposed as the difference of two test field $y_{ab}$ and $z_{ab}$, i.e., $x_{ab} = z_{ab} - y_{ab}$, both satisfying the dual energy condition everywhere.  In that case, we have,
\[
\int_M T^{ab}x_{ab} = \int_M T^{ab} z_{ab} - \int_M T^{ab} y_{ab},
\]
which, since both integrals on the right hand side are always non-negative, yields a positive number if and only if there is ``more'' $T^{ab}$ in the region of support of $y_{ab}$ (by the standard given by $y_{ab}$) than there is in the region of support of $z_{ab}$ (again, by the standard given by $z_{ab}$).  (See Fig. \ref{fig:tracking}.)

With these remarks in hand, we return to the definition of tracking.  There we require that, for \emph{any} test field $x_{ab}$, satisfying the dual energy condition in a neighborhood of $\gamma$, there is an element of $\mathcal{C}$ whose action on $x_{ab}$ is (strictly) positive.  This captures the idea that, by any standard one likes---or at least, any standard captured by test fields---for measuring ``amount of $T^{ab}$ near $\gamma$'' and ``amount of $T^{ab}$ away from $\gamma$'', $T^{ab}$ may be chosen from $\mathcal{C}$ so that there is more $T^{ab}$ near $\gamma$ than away from $\gamma$---or in other words, there exist fields $T^{ab}$ in $\mathcal{C}$ that are as concentrated as one wishes near $\gamma$.   Note that we consider only those ``standards of measurement'' given by test fields, which always have compact support.  This means that we are considering $T^{ab}$ fields that are arbitrarily concentrated near $\gamma$ for arbitrarily long, but finite, duration.  It also means that there could be arbitrarily large amounts of $T^{ab}$ far from $\gamma$, as long as it does not fall within the support of $x_{ab}$.

\subsection{Consequences of tracking}

As we have just seen, tracking gives us a sense in which a collection of fields includes elements that follow a curve $\gamma$ as closely as one likes, for as long as one likes.  It then follows that a collection $\mathcal{C}$, satisfying various properties, can track only certain curves.  In particular, we get the following result.\footnote{These results rephrase Theorem 3 and the subsequent discussion  of \citet{Geroch+Weatherall}.}
\begin{thm}\label{tt} Let $(M,g_{ab})$ be a spacetime, $\gamma$ a timelike curve therein, and ${\cal C}$ a collection of symmetric fields $T^{ab}$, each satisfying the dominant energy condition, that tracks $\gamma$.   Suppose each of these fields is conserved.  Then there exists a sequence of fields $T_1^{ab}, T_2^{ab}, \ldots$, each a positive multiple of some element of ${\cal C}$, that converges, in the sense of distributions, to ${\boldsymbol\delta}_{\gamma}u^a u^b$.
\end{thm}
\begin{cor}\label{ttcor}
The curve $\gamma$ is a geodesic.
\end{cor}
With a small modification, Theorem \ref{tt} and Corollary \ref{ttcor} hold for null curves as well.\footnote{The small modification involves the definition of a $\delta$ distribution supported on a null curve, which requires a choice of parameterization (since null curves cannot be parameterized by arc-length).  It does not affect the conclusion that $\gamma$ is a geodesic.}  We also have the following converse:\footnote{Observe that this converse may be understood to capture a sense in which superluminal propagation is impossible, at least in a point-particle limit.  One might take this result to be in tension with the arguments of \citet{GerochSL} and \citet{WeatherallSL}.  But in fact, the tension is only apparent: this result assumes the dominant energy condition, while the discussions in those other papers do not (see also \citet{EarmanSL} for a discussion of the relation between the dominant energy condition and the notion of ``superluminal propagation'' discussed there).  That said, the present result, in connection with those earlier papers, raises an interesting question.  Can one generalize the notion of tracking considered here to hyperbolic systems whose solutions do not satisfy the dominant energy condition, and if so, do solutions always track their characteristics?  I am grateful to an anonymous referee for raising the possible tension.}
\begin{thm}\label{slc}Let $(M,g_{ab})$ be a spacetime, $\gamma$ a curve therein, and ${\cal C}$ a collection of symmetric fields $T^{ab}$, each satisfying the dominant energy condition that tracks $\gamma$.  Then $\gamma$ is timelike or null.
\end{thm}

These results imply that the only curves that collections of smooth, symmetric, rank 2 fields, all divergence-free and satisfying the dominant energy condition, can track are timelike or null geodesics.  But they also say more than this: they assert that \emph{any} family of bodies, satisfying the energy condition and the conservation condition, that follows a curve arbitrarily tightly contains a sequence converging, up to rescaling, to the (distributional) energy-momentum representing a point particle.   In other words, \emph{every} sequence of smooth, symmetric, divergence-free fields, satisfying the dominant energy condition, whose support approaches a timelike or null curve $\gamma$, converges, up to rescaling, to a multiple of the $\delta$ distribution on $\gamma$.  This captures the sense in which the distribution ${\boldsymbol\delta}_{\gamma}u^au^b$ represents the energy-momentum of realistic (extended) matter: it is the essentially unique accumulation point for energy-momentum tensors of small bodies.  The key insight is that bodies that can be made arbitrarily small in size, in the sense captured by tracking, necessarily approach delta functions on a curve.\footnote{\label{spin2} Consider this result in connection with the remarks in footnote \ref{spin1}: as noted there, the dominant energy condition for distributions is incompatible with higher order distributions, and thus, with point particles carrying non-vanishing angular momentum.  Here we see an even stronger result, which is that, in the small body limit, extended bodies all satisfying the dominant energy condition must have vanishing angular momentum (per unit mass).}

So Theorem \ref{tt} makes direct contact with the distributional results described in section \ref{sec:two}, and it clarifies the physical significance of order-zero, divergence-free distributions $\mathbf{T}^{ab}$ supported on a curve.  But this theorem \emph{also} captures, and indeed strengthens, the Geroch-Jang result.  To see this, observe the following:
\begin{prop}If ${\cal C}$ contains, for every neighborhood $O$ of a curve $\gamma$, a smooth, symmetric, non-vanishing, divergence-free field $T^{ab}$ that satisfies the dominant energy condition and vanishes outside of $O$, then ${\cal C}$ tracks $\gamma$.\end{prop}
In other words, the antecedent of the Geroch-Jang theorem implicitly defines a collection $\mathcal{C}$ of fields $T^{ab}$, each satisfying the dominant energy condition and each divergence-free: these are the fields that are supported (only) in arbitrarily small neighborhoods of a curve $\gamma$.  What this proposition asserts is that this collection tracks $\gamma$; it follows, then, from Theorem \ref{tt} that not only is $\gamma$ a geodesic, but that the collection $\mathcal{C}$ contains a sequence that, up to rescaling, converges to a $\delta$ distribution on $\gamma$.  Moreover, we see how the collection $\mathcal{C}$ defined by the Geroch-Jang theorem is more restrictive than necessary to get this result---and thus we see the sense in which Theorem \ref{tt} is a strict strengthening of the Geroch-Jang theorem.  One can recover the Ehlers-Geroch theorem in a similar manner.

\subsection{Applications of these results}

The results just described allow us to extend the curve-first approach in two important ways.  First, by connecting curve-first and distributional results, Theorem \ref{tt} provides an important hint on how to extend the curve-first approach to forced motion.  In particular, we see that well-chosen collections $\mathcal{C}$ that track curves accumulate, up to re-scaling, on unique distributions on a curve.  Thus, to get curve-first results for forced motion, we need to exert enough control on the collection ${\cal C}$ to specify a limit up to overall scaling.  And to see how to exercise that control, we can investigate the character of the distributional results.

For instance, in the case of a charged body, requiring a distributional $\mathbf{T}^{ab}$, supported on a curve $\gamma$, to satisfy the dominant energy condition implies that the charge-current density must be, at most, order one.  As noted in section \ref{sec:two}, one can give a complete treatment of this case; when one does so, one finds contributions to the motion of the body arising from dipole moments of the charge-current density.  Our only hope of getting a unique distributional limit, then, is if we can somehow control, on the way to the limit, what the contributions from the electric and magnetic dipole moments will be---for instance, by requiring that they be suitably bounded, in the limit, by the mass of the body.

We make this idea precise as follows.  Let ${\cal C}$ be a collection of pairs $(T^{ab}, J^a)$ of smooth fields, where each $T^{ab}$ satisfies the dominant energy condition.  We will say that a number $\kappa > 0$ bounds the charge-to-mass ratio of the elements of ${\cal C}$ if, for any unit timelike vector $t^a$ at a point, and any pair $(T^{ab}, J^a)\in{\cal C}$, \[|J^a t_a| \leq \kappa T^{ab}t_at_b.\]  This condition captures the idea that the collection $\mathcal{C}$ does not include elements whose charge density relative to any observer becomes arbitrarily large, relative to its mass density.  Since in the small body limit, a ``dipole moment'' looks like a charge density that goes from infinitely large (and positive) to infinitely large (and negative) over a vanishingly small region, bounding the charge density in this way forces contributions from dipole moments to vanish in the small body limit.  On a more technical level, since we know that $T^{ab}$ exhibits order zero behavior in the small body limit (by virtue of the energy condition), bounding $J^a$ by $T^{ab}$ enforces order zero behavior on $J^a$ as well.

We then get the following result.
\begin{thm}\label{Lorentz} Let $(M, g_{ab})$ be a space-time, $F_{ab}$ an antisymmetric tensor field on $M$, and $\gamma$ a timelike curve.  Let ${\cal C}$ be a collection of pairs, $(T^{ab}, J^a)$, of tensor fields on $M$, where each $T^{ab}$ satisfies the dominant energy condition, each $J^a$ satisfies $\nabla_a J^a = 0$, and each pair satisfies $\nabla_b T^{ab} = F^a{}_b J^b$.  Suppose the collection has charge-mass ratio bounded by $\kappa \geq 0$ and that it tracks $\gamma$.  Then there exists a sequence of pairs, $(\overset{n}{T}{}^{ab}, \overset{n}{J}{}^a)$, each a multiple of some element of ${\cal C}$, that converges, as distributions, to $(u^a u^b \mathbf{\delta}_\gamma, \kappa' u^a \mathbf{\delta}_\gamma)$, for some number $\kappa'$ satisfying $|\kappa'| \leq \kappa$.\end{thm}
\begin{cor}
The curve $\gamma$ is a Lorentz force curve with charge-to-mass ratio $\kappa'$.
\end{cor}
This result captures a sense in which the Lorentz force law is a theorem of electromagnetism---and it also shows that this theorem has the same ``curve-first'' character as, say, the Geroch-Jang theorem.  Note that a crucial assumption in Theorem \ref{Lorentz} is that for each pair $(T^{ab}, J^a)$ in $\mathcal{C}$, $\nabla_b T^{ab} = F^{a}{}_{b} J^b$---just as, in Theorem \ref{tt}, a crucial assumption is that $\nabla_b T^{ab}=\mathbf{0}$.\footnote{Note, too, that the subtleties regarding that status of the conservation condition discussed in \citet{WeatherallSGP,WeatherallConservation} arise here, too: in particular, although for \emph{sources} to Maxwell's equation, $\nabla_bT^{ab} = F^a{}_b J^b$ holds automatically, as a consequence of Maxwell's equations (just as $\nabla_b T^{ab}=\mathbf{0}$ holds for sources in Einstein's equation), here we are considering \emph{test} matter in Maxwell's equations, since the background field $F_{ab}$ is fixed in advance. One could imagine considering a variation of this result, along the lines of the Ehlers-Geroch theorem, that allows electromagnetic backreaction, or even that allows both electromagnetic and gravitational backreaction.  Though I do not know of any technical barriers to such results, formulating them is a delicate matter and we have not pursued it.}

So we see that Theorem \ref{tt} and it corollaries substantially strengthen the consequent of curve-first results---by giving the universal limiting behavior of certain sequences of smooth fields---and in doing so, provides hints about how to extend these results to forced motion.  This is one way in which they extend curve-first results.  The second way is that they weaken the premises.  In particular, they permit matter to be non-vanishing far from $\gamma$, as long as the quantity of such matter can be made arbitrarily small in any particular region. Hence, these results apply to solutions of hyperbolic systems, such as Maxwell's equations and the Klein-Gordon equation.  The basic idea is that the solutions to a hyperbolic system---say, Maxwell's equations---naturally give rise to a collection $\mathcal{C}$ of smooth fields $T^{ab}$. Insofar as these collections satisfy the dominant energy condition and are divergence-free, we can then apply the theorems above.

More precisely, fix a globally hyperbolic spacetime $(M,g_{ab})$, and let ${\cal C}$ be the collection of energy-momentum tensors associated with solutions of the source-free Maxwell equations on that spacetime.\footnote{We require that the spacetime be globally hyperbolic so that we are certain to have ``enough'' solutions to Maxwell's equations; one could imagine relaxing this requirement.}  It immediately follows, as a consequence of Maxwell's equations themselves, that each element of ${\cal C}$ is divergence-free and satisfies the dominant energy condition.\footnote{See \citet[\S 2.6]{MalamentGR} for a discussion of this point.} We can thus apply Theorem \ref{slc} to conclude that ${\cal C}$ can track \emph{only} timelike and null curves; and apply Theorem \ref{tt} to conclude that if it tracks any curves at all, they must be geodesics.

We cannot, however, conclude from the general analysis that $\mathcal{C}$ tracks any curves at all.  For that, we need to analyze the solutions to Maxwell's equations.\footnote{The relevant arguments concerning Maxwell's equations, and the other equations discussed below, are given in \citet[\S 4]{Geroch+Weatherall}.}  In fact, we find that ${\cal C}$ tracks all null geodesics; it tracks no timelike geodesics.  It follows that there exist sequences of electromagnetic fields whose energy-momentum tensors converge to multiples of a $\delta$ distribution supported on null geodesics. This captures the sense in which light rays follow null geodesics, and it makes the so-called ``optical limit'' of electromagnetism a special case of more general theorems concerning small-body motion.  Note, however, that we have not avoided the sort of reasoning that goes into the optical limit altogether---the fact that one can form long-lasting wave packets with high frequency solutions to Maxwell's equations is essential to the argument that $\mathcal{C}$ tracks any curves at all.

It is important to emphasize how this approach has avoided the problem with distributional solutions to hyperbolic systems described in section \ref{sec:two}: we do not require the electromagnetic fields themselves to converge to any distribution, and so we do not claim that the limiting distribution $\mathbf{T}^{ab}$ is the energy-momentum distribution associated with any particular solution.  Rather, we claim that the limiting distribution approximates the energy-momentum properties of real solutions that are concentrated near a curve, without having any ``underlying'' field associated with it.

I will conclude this section by briefly discussing one more example, because it has some unexpected features.  Consider the collection ${\cal C}$ of energy-momentum tensors associated with solutions of the mass $m$ Klein-Gordon equation on our spacetime $(M,g_{ab})$.  As with Maxwell's equations, each element of ${\cal C}$ is divergence-free and satisfies the dominant energy condition, and so once again $\mathcal{C}$ can track \emph{only} timelike and null geodesics.  In fact, one can show that ${\cal C}$ tracks all null geodesics; it tracks no timelike geodesics.

This result is perhaps surprising: after all, one might expect mass $m>0$ Klein-Gordon fields to be \emph{massive}, i.e., to give rise, in the small-body limit, to massive particles, following timelike, not null, geodesics.  The reason this does not happen turns on an ambiguity in the meaning of ``mass''.  The parameter $m$ in the Klein-Gordon equation does, in a certain sense, characterize the mass of the particle.  But given a solution to the Klein-Gordon equation, $m$ is neither the mass density associated with the solution at any point, nor is it the ``total mass'' associated with any spacelike slice (if suitable slices even exist).\footnote{In what follows, when I write of ``total mass'', readers who are troubled by this notion should suppose we are in Minkowski spacetime, or a suitable asymptotically flat spacetime, where such notions make sense.}  And if we are thinking of the ``particle'' that arises in the small body limit of solutions to the Klein-Gordon equation, we should generally expect, for \emph{any} fixed mass $m>0$, that as the spatial support of the body approaches a curve, the ``total mass'' of the body, i.e., the integrated mass on a suitable spacelike slice, will approach zero.  Thus, for any fixed $m$, we should think of small-body Klein-Gordon solutions as behaving like massless particles.  Another way to see the same conclusion is that, if one imagines trying to make a Klein-Gordon wave packet propagate more and more tightly along a curve, one needs to move to higher and higher frequency solutions.  But these correspond to higher and higher velocities for the ``massive'' particle one is trying to construct, and ultimately converge to a null geodesic.

If we want to consider particles that are ``massive'', even in the limit, then, we need to consider not solutions the Klein-Gordon equation for fixed $m$, but rather solutions of the mass $m$ Klein-Gordon equation \emph{for all $m> 0$}.  With this modification, we find that the collection ${\cal C}$ of energy-momentum tensors associated with all such solutions tracks all timelike and null geodesics.

Finally, I remark that one can also consider charged Klein-Gordon fields with a fixed background electromagnetic field; in this case, one can construct a collection $\mathcal{C}$ of pairs of energy-momentum tensors and charge-current densities for all solutions to Klein-Gordon equations with $m> 0$ and fixed charge-to-mass ratio $\kappa$.  This collection will satisfy the conditions of Theorem \ref{Lorentz}, and so these fields can track only Lorentz force curves (and null curves).

\section{Dynamics, Inertia, and Spacetime Geometry}

As I noted in the introduction, inertial structure, encapsulated by the geodesic principle, provides a powerful link between motion and physical geometry.  It identifies a geometrically privileged class of curves with a physically  privileged class of motions---hence giving physical significance to the notion of ``geodesy''.  This result also has a converse, which I did not mention above: all metric geometry is encoded in the class of inertial trajectories.  In particular, a classic result due to \citet{Weyl} establishes that if two Lorentzian metrics agree on all null and timelike geodesics, up to reparameterization, then they are constant multiples of one another \citep[Prop. 2.1.4]{MalamentGR}.

But the geodesic principle concerns point particles, and as I argued in section \ref{sec:two}, the status of such objects is unclear in general relativity.  This puts some pressure on the foundational significance of the geodesic principle---and on the link between geometry and motion that it provides.  What should we make of a foundational principle that, by the lights of the theory of which it is part, relies on the counterfactual behavior of impossible objects?\footnote{One might respond that the Geroch-Jang and Ehlers-Geroch theorems do not explicitly refer to point particles, and so these, too, permit one to state the geodesic principle without reference to point particles.  Fair enough.  But from my perspective the main appeal of the current proposal is precisely that it is an assertion \emph{about field equations}, and as we have seen, this is precisely what one cannot get from the Geroch-Jang and Ehlers-Geroch constructions.  I am grateful to an anonymous referee for raising this objection.}

Fortunately, there is another way.  The methods described in section \ref{sec:tracking} provide the resources to capture the link between motion and physical geometry directly via the solutions to matter field equations (i.e., hyperbolic systems), without any reference to point particles.  The key idea is, once again, tracking, which allows us to state a new form of the geodesic principle as follows: The energy-momentum tensors associated with solutions to source-free matter field equations track (only) timelike or null geodesics.

What does this formulation express? First, it once again captures something about inertial, i.e., force-free, motion.  This is because we restrict attention to \emph{source-free} fields, where we understand sources to be interactions with other forms of matter.\footnote{There is an interesting question lurking in the background here, which is: can we always unambiguously identify ``source terms'' in a differential equation?  In standard cases in physics, it is generally clear what counts as a source.  But I will not attempt to give an analysis of this concept here, and will proceed on the assumption that it is sufficiently clear for current purposes.}   It is these solutions that one would expect to be associated with divergence-free energy-momentum tensors.  It also associates certain force-free motions of physical bodies with a geometrically privileged class of curves.  Now, though, that association runs via a particular limiting construction, concerning the curves near which solutions to these equations can be made to propagate.  It tells us something about how the solutions to these equations behave.

Remarkably, in this new form the geodesic principle  is (almost) a theorem \emph{as stated}.  The results in section \ref{sec:tracking} establish that it holds for a system of field equations whenever the energy-momentum tensors associated with source-free solutions have two properties: (1) they are divergence-free with respect to the spacetime derviative operator $\nabla$; and (2) they satisfy the dominant energy condition.

The first of these conditions holds in considerable generality for matter whose dynamics can be derived from a Lagrangian density in a certain standard way.\footnote{This claim is well-known and widely discussed in the physics literature; see for instance \citet[Appendix E]{Wald} for an argument. For further discussion in a foundational context, with particular emphasis on the relationship between this claim and the geodesic principle in general relativity and other theories, see \citet{WeatherallConservation}.}  In particular, consider some species of matter $\Phi^X$ in a spacetime $(M,g_{ab})$. Suppose the dynamics for $\Phi^X$ follow from extremizing an action $I[\Phi^X,g_{ab}]=\int_M \mathcal{L}(\Psi^X,g_{ab})$ depending {only} on $g_{ab}$, $\Psi^X$, and its covariant derivatives.  If $\Phi^X_0$ is a solution to the resulting equations in $(M,g_{ab})$, then $T^{ab}:=\left(\frac{\delta \mathcal{L}}{\delta g_{ab}}\right)_{|\Psi^X_0, g_{ab}}$ is divergence-free with respect to the derivative operator compatible with $g_{ab}$.  Hence, for a broad class of matter that includes all candidates for ``fundamental'' matter fields in general relativity, energy-momentum is conserved relative to the metric appearing in its dynamics---i.e., the one determining the notions of length, duration, and angle salient to its evolution.

But what about the energy condition?\footnote{For a discussion of the status of energy conditions in general relativity, see \citet{CurielEC}.}  First, we remark that an energy condition is essential to the arguments given in section \ref{sec:tracking}.  In particular, the dominant energy condition plays two roles there.  First, it enforces ``positivity''.  The basic idea behind tracking is to use the fact that we have a class of tensor fields whose action on a certain class of test fields is always non-negative, to ``measure'' the energy-momentum in different regions.  This, recall, is how we capture the idea that there is ``more'' $T^{ab}$ in a region near a curve than there is far from the curve.  If we tried to drop the energy condition all together, tracking would no longer make sense, because the fields $T^{ab}$ under consideration would not necessarily be positive when acting on any particular set of test fields.  On the other hand, it is likely that a weaker energy condition would suffice in this role.  The key seems to be to require that all fields $T^{ab}$ in a collection $\mathcal{C}$ lie, at each point, within some convex cone.

The second role that the energy condition plays is that it enforces causality.  That is, the dominant energy condition is what rules out the possibility of collections tracking \emph{spacelike} geodesics, as in Theorem \ref{slc}.  It does not appear to be the case that weaker energy conditions could suffice for this role.\footnote{In effect, this is what is shown in \citet{WeatherallEC}.  Note, however, that the \emph{strengthened} dominant energy condition considered there, which is necessary for the Geroch-Jang theorem as stated, would not be natural in the current context.  The reason is that distributions do not take values at points, and so requiring that they have certain behavior at points where they are non-vanishing is awkward to express.  At best one would have to recast the condition in terms of the support of the distribution.}  Thus, it seems to be the case that one could relax the energy condition, and still conclude that solutions to a field equation track only geodesics.  But the full claim that a collection of $T^{ab}$ fields tracks only timelike or null geodesics apparently requires at least the dominant energy condition.

So we need the dominant energy condition.  Fortunately, it holds, automatically, for many fields of physical interest.  For instance, the dominant energy condition always holds for the energy-momentum tensors associated with source-free solutions to Maxwell's equations, for solutions to the (non-negative mass) Klein-Gordon equation, and so on.  But it does not hold for \emph{all} equations that one might be interested in.  In particular, solutions to the Dirac equation may not satisfy even the weak energy condition.\footnote{Observe that this failure to satisfy the energy conditions is not obviously related to the fact that Dirac fields have ``intrinsic'' angular momentum (though it is related to the fact that they are spinors).  (Recall fns. \ref{spin1} and \ref{spin2}.)}  This suggests that it is the dominant energy condition that is key to whether a given form of matter, with dynamics derivable from a suitable Lagrangian, will satisfy the (new) geodesic principle---and also that it is not clear that all matter fields of physical interest \emph{do} satisfy the new geodesic principle.\footnote{One might worry that this last observation is a problem for the proposed formulation of the geodesic principle in terms of tracking.  But I do not think there is a real concern.  Source-free matter that tracks non-geodesic curves is every bit as much a problem for other formulations of the geodesic principle as the present one---and at least on the proposed formulation, the tension between such matter and geodesic motion is immediately manifest.}

This discussion suggests that the status of the dominant energy condition deserves more attention.  In particular, one would like to identify the conditions under which the energy-momentum tensor associated with solutions to a given matter field equation are certain to satisfy the energy condition.  Of special interest would be to articulate the relationship between the dominant energy condition, on the one hand, and the ``causal cone'' associated with a hyperbolic system, which captures a (different) sense in which solutions to a system of equations may propagate causally.\footnote{There has been some discussion of this relationship in both the physics and philosophy literatures \citep{GerochPDE,EarmanSL,WeatherallSL,Wong}, but it does not seem that a fully satisfactory answer is available.}

\section*{Acknowledgments}
I am grateful to Harvey Brown, Bob Geroch, Dennis Lehmkuhl, David Malament, and Bob Wald for many helpful conversations related to this material.  I am also grateful to audiences at the Eighth Quadrennial International Fellows Conference (2016) in Lund, Sweden; New Directions in Philosophy of Physics (2017) in Tarquinia, Italy; the Pacific Institute of Theoretical Physics at the University of British Columbia; the workshop ``Thinking about Space and Time'' at the University of Bern; the Max Planck Institute for Gravitational Physics in Potsdam, Germany; the Black Hole Initiative at Harvard University; the Einstein Papers Project at Caltech; and Utrecht University, and especially to Lars Andersson, Joshua Goldberg, Jim Hartle, Bob Wald, and Shin-Tung Yau for probing questions, comments, and objections.

\singlespacing

\bibliographystyle{elsarticle-harv}
\bibliography{motion}

\begin{thebibliography}{41}
\expandafter\ifx\csname natexlab\endcsname\relax\def\natexlab#1{#1}\fi
\expandafter\ifx\csname url\endcsname\relax
  \def\url#1{\texttt{#1}}\fi
\expandafter\ifx\csname urlprefix\endcsname\relax\def\urlprefix{URL }\fi

\bibitem[{Asada et~al.(2011)Asada, Futamase, and Hogan}]{Asada}
Asada, H., Futamase, T., Hogan, P.~A., 2011. Equations of motion in general
  relativity. Oxford University Press, Oxford, UK.

\bibitem[{Brown(2005)}]{Brown}
Brown, H.~R., 2005. Physical Relativity. Oxford University Press, New York.

\bibitem[{Colombeau(2000)}]{Colombeau}
Colombeau, J.~F., 2000. New generalized functions and multiplication of
  distributions. Vol.~84. Elsevier.

\bibitem[{Curiel(2017)}]{CurielEC}
Curiel, E., 2017. A primer on energy conditions. In: Lehmkuhl, D., Schiemann,
  G., Scholz, E. (Eds.), Towards a Theory of Spacetime Theories. Birkh\"auser,
  Boston, MA, pp. 43--104.

\bibitem[{D'Eath(1975)}]{Death}
D'Eath, P.~D., 1975. Interaction of two black holes in the slow-motion limit.
  Physical Review D 12~(8), 2183.

\bibitem[{Earman(2014)}]{EarmanSL}
Earman, J., 2014. No superluminal propagation for classical relativistic and
  relativistic quantum fields. Studies in History and Philosophy of Science
  Part B: Studies in History and Philosophy of Modern Physics 48, 102--108.

\bibitem[{Ehlers and Geroch(2004)}]{Ehlers+Geroch}
Ehlers, J., Geroch, R., 2004. Equation of motion of small bodies in relativity.
  Annals of Physics 309, 232--236.

\bibitem[{Einstein and Grommer(1927)}]{Einstein+Grommer}
Einstein, A., Grommer, J., 1927. Allgemeine Relativit{\"a}tstheorie und
  Bewegungsgesetz. Verlag der Akademie der Wissenschaften, Berlin.

\bibitem[{Geroch(1996)}]{GerochPDE}
Geroch, R., 1996. Partial differential equations of physics,
  arXiv:gr-qc/9602055.

\bibitem[{Geroch(2011)}]{GerochSL}
Geroch, R., 2011. Faster than light? In: Plaue, M., Rendall, A., Scherfner, M.
  (Eds.), Advances in Lorentzian Geometry. American Mathematical Society,
  Providence, RI, pp. 59--80.

\bibitem[{Geroch and Jang(1975)}]{Geroch+Jang}
Geroch, R., Jang, P.~S., 1975. Motion of a body in general relativity. Journal
  of Mathematical Physics 16~(1), 65.

\bibitem[{Geroch and Traschen(1987)}]{Geroch+Traschen}
Geroch, R., Traschen, J., 1987. Strings and other distributional sources in
  general relativity. Physical Review D 36~(4), 1017.

\bibitem[{Geroch and Weatherall(2018)}]{Geroch+Weatherall}
Geroch, R., Weatherall, J.~O., 2018. The motion of small bodies in space-time.
  Communications in Mathematical PhysicsForthcoming.
  DOI:10.1007/s00220-018-3268-8.

\bibitem[{Gralla et~al.(2009)Gralla, Harte, and Wald}]{Harte}
Gralla, S.~E., Harte, A.~I., Wald, R.~M., 2009. Rigorous derivation of
  electromagnetic self-force. Physical Review D 80~(2), 024031.

\bibitem[{Gralla and Wald(2011)}]{Wald+Gralla}
Gralla, S.~E., Wald, R.~M., 2011. A rigorous derivation of gravitational
  self-force. Classical and Quantum Gravity 28~(15), 159501.

\bibitem[{Grosser et~al.(2001)Grosser, Kunzinger, Oberguggenberger, and
  Steinbauer}]{distributionGR}
Grosser, M., Kunzinger, M., Oberguggenberger, M., Steinbauer, R., 2001.
  Geometric theory of generalized functions with applications to general
  relativity. Vol. 537. Springer Science \& Business Media, Dordrecht.

\bibitem[{Lehmkuhl(2017{\natexlab{a}})}]{LehmkuhlHybrid}
Lehmkuhl, D., 2017{\natexlab{a}}. General relativity as a hybrid theory: The
  genesis of {E}instein’s work on the problem of motion. Studies in History
  and Philosophy of Modern PhysicsForthcoming. DOI:10.1016/j.shpsb.2017.09.006.

\bibitem[{Lehmkuhl(2017{\natexlab{b}})}]{LehmkuhlCareful}
Lehmkuhl, D., 2017{\natexlab{b}}. Literal versus careful interpretations of
  scientific theories: The vacuum approach to the problem of motion in general
  relativity. Philosophy of Science 84~(5), 1202--1214.

\bibitem[{Malament(2012{\natexlab{a}})}]{MalamentGP}
Malament, D., 2012{\natexlab{a}}. A remark about the ``geodesic principle'' in
  general relativity. In: Frappier, M., Brown, D.~H., DiSalle, R. (Eds.),
  Analysis and Interpretation in the Exact Sciences: Essays in Honour of
  William Demopoulos. Springer, New York, pp. 245--252.

\bibitem[{Malament(2012{\natexlab{b}})}]{MalamentGR}
Malament, D.~B., 2012{\natexlab{b}}. Topics in the Foundations of General
  Relativity and Newtonian Gravitation Theory. University of Chicago Press,
  Chicago.

\bibitem[{Mathisson(1931)}]{Matthisson}
Mathisson, M., 1931. Die mechanik des materieteilchens in der allgemeinen
  relativit\"atstheorie''. Zeitschrift f\"ur Physik 67, 826--844.

\bibitem[{Mino et~al.(1997)Mino, Sasaki, and Tanaka}]{Mino+etal}
Mino, Y., Sasaki, M., Tanaka, T., 1997. Gravitational radiation reaction to a
  particle motion. Physical Review D 55~(6), 3457.

\bibitem[{Papapetrou(1951)}]{Papapetrou}
Papapetrou, A., 1951. Spinning test-particles in general relativity. i. Proc.
  R. Soc. Lond. A 209~(1097), 248--258.

\bibitem[{Poisson et~al.(2011)Poisson, Pound, and Vega}]{Poisson}
Poisson, E., Pound, A., Vega, I., 2011. The motion of point particles in curved
  spacetime. Living Reviews in Relativity 14~(7).

\bibitem[{Puetzfeld et~al.(2015)Puetzfeld, L{\"a}mmerzahl, and
  Schutz}]{Puetzfeld}
Puetzfeld, D., L{\"a}mmerzahl, C., Schutz, B. (Eds.), 2015. Equations of motion
  in relativistic gravity. Springer, Heidelberg, Germany.

\bibitem[{Samaroo(2015)}]{Samaroo}
Samaroo, R., 2015. There is no conspiracy of inertia. British Journal for the
  Philosophy of ScienceForthcoming.

\bibitem[{Souriau(1974)}]{Souriau}
Souriau, J.-M., 1974. Mod\`ele de particule \`a spin dans le champ
  \'electromagn\'etique et gravitationnel. Annales de l'Institut Henri
  Poincar\'e Sec. A 20, 315.

\bibitem[{Steinbauer and Vickers(2006)}]{Steinbauer+Vickers}
Steinbauer, R., Vickers, J.~A., 2006. The use of generalized functions and
  distributions in general relativity. Classical and Quantum Gravity 23~(10),
  R91.

\bibitem[{Sternberg and Guillemin(1984)}]{Sternberg+Guillemin}
Sternberg, S., Guillemin, V., 1984. Symplectic Techniques in Physics. Cambridge
  University Press, Cambridge.

\bibitem[{Sus(2014)}]{Sus}
Sus, A., 2014. On the explanation of inertia. Journal for General Philosophy of
  Science 45~(2), 293--315.

\bibitem[{Tamir(2012)}]{Tamir}
Tamir, M., 2012. Proving the principle: Taking geodesic dynamics too seriously
  in {E}instein's theory. Studies in History and Philosophy of Modern Physics
  43~(2), 137--154.

\bibitem[{Thorne and Hartle(1985)}]{Thorne+Hartle}
Thorne, K.~S., Hartle, J.~B., 1985. Laws of motion and precession for black
  holes and other bodies. Physical Review D 31~(8), 1815.

\bibitem[{Wald(1984)}]{Wald}
Wald, R.~M., 1984. General Relativity. University of Chicago Press, Chicago.

\bibitem[{Weatherall(2011{\natexlab{a}})}]{WeatherallMBNT}
Weatherall, J.~O., 2011{\natexlab{a}}. The motion of a body in {N}ewtonian
  theories. Journal of Mathematical Physics 52~(3), 032502.

\bibitem[{Weatherall(2011{\natexlab{b}})}]{WeatherallSGP}
Weatherall, J.~O., 2011{\natexlab{b}}. On the status of the geodesic principle
  in {N}ewtonian and relativistic physics. Studies in the History and
  Philosophy of Modern Physics 42~(4), 276--281.

\bibitem[{Weatherall(2012)}]{WeatherallEC}
Weatherall, J.~O., 2012. A brief remark on energy conditions and the
  {G}eroch-{J}ang theorem. Foundations of Physics 42~(2), 209--214.

\bibitem[{Weatherall(2014)}]{WeatherallSL}
Weatherall, J.~O., 2014. Against dogma: On superluminal propagation in
  classical electromagnetism. Studies in History and Philosophy of Science Part
  B: Studies in History and Philosophy of Modern Physics 48, 109--123.

\bibitem[{Weatherall(2017{\natexlab{a}})}]{WeatherallConservation}
Weatherall, J.~O., 2017{\natexlab{a}}. Conservation, inertia, and spacetime
  geometry. Studies in History and Philosophy of Science Part B: Studies in
  History and Philosophy of Modern PhysicsForthcoming.

\bibitem[{Weatherall(2017{\natexlab{b}})}]{WeatherallLehmkuhl}
Weatherall, J.~O., 2017{\natexlab{b}}. Inertial motion, explanation, and the
  foundations of classical spacetime theories. In: Lehmkuhl, D., Schiemann, G.,
  Scholz, E. (Eds.), Towards a Theory of Spacetime Theories. Birkh\"auser,
  Boston, MA, pp. 13--42.

\bibitem[{Weyl(1922)}]{Weyl}
Weyl, H., 1922. Space--Time--Matter. Methuen \& Co., London, UK, reprinted in
  1952 by Dover Publications.

\bibitem[{Wong(2011)}]{Wong}
Wong, W. W.-Y., 2011. Regular hyperbolicity, dominant energy condition and
  causality for lagrangian theories of maps. Classical and Quantum Gravity
  28~(21), 215008.

\end{thebibliography}

\end{document}